\documentclass[twocolumn,aps,prb,showpacs]{revtex4}

\usepackage{epsfig}

\begin{document}

\title{Spin susceptibility in bilayered cuprates:
resonant magnetic excitations}
\author{Ilya Eremin$^{1}$, Dirk K. Morr$^{2,3}$, Andrey V. Chubukov$^4$,
Karl Bennemann$^2$}
 \affiliation{$^1$ Max-Planck Institut f\"ur
Physik komplexer Systeme, D-01187 Dresden, Germany and Institute
f\"ur Mathematische und Theoretische Physik, TU-Braunschweig,
D-38106 Braunschweig, Germany \\$^2$ Institute f\"ur Theoretische
Physik, Freie Universit\"{a}t Berlin, D-14195, Berlin, Germany \\
$^3$ Department of Physics, University of Illinois at Chicago,
Chicago, IL 60607\\ $^4$ Department of Physics, University of
Wisconsin - Madison, Madison, WI 53706}
\date{\today}

\begin{abstract}
We study the momentum and frequency dependence of the dynamical spin
susceptibility in the superconducting state of bilayer cuprate
superconductors. We show that there exists a resonance mode in the
odd as well as the even channel of the spin susceptibility, with the
even mode being located at higher energies than the odd mode. We
demonstrate that this energy splitting between the two modes arises
not only from a difference in the interaction, but also from a
difference in the free-fermion susceptibilities of the even and odd
channels. Moreover, we show that the even resonance mode disperses
downwards at deviations from ${\bf Q}=(\pi,\pi)$. In addition, we
demonstrate that there exists a second branch of the even resonance,
similar to the recently observed second branch (the $Q^*$-mode) of
the odd resonance. Finally, we identify the origin of the
qualitatively different doping dependence of the even and odd
resonance. Our results suggest further experimental test that may
finally resolve the long-standing question regarding the origin of
the resonance peak.

\end{abstract}

 \maketitle

\section{Introduction}

Magnetic excitations in the high-temperature superconductors are of
fundamental interest. While it is currently still a topic of intense
debate whether a continuum of magnetic excitations is responsible
for the occurrence of superconductivity in the cuprates, the
feedback effect of $d_{x^2-y^2}$-wave superconductivity on the
magnetic excitation spectrum has been well established in the
context of the ``resonance peak". This peak has been observed by
inelastic neutron scattering (INS) experiments in three different
families of the high-temperature superconductors \cite{YBCO,Bi,Tl}.
The doping dependence of the peak frequency, $\Omega_{res} ({\bf
Q})$, the downward dispersion of the resonance, which tracks the
momentum dependence of the particle-hole continuum, and the
emergence of a second resonance branch further away from ${\bf Q}$
are all consistent with the idea that the resonance peak is a
particle-hole bound state (i.e. a {\it spin exciton}) below the
particle-hole continuum. According to theory~\cite{excit}, this
excitonic resonance is a fundamental property of a $d_{x^2-y^2}$
superconductor. (For a review of other theoretical scenarios, see
Refs.~\cite{other,other1,other2}).

Recent INS experiments in overdoped YBa$_2$Cu$_3$O$_{6+x}$ (YBCO)
revealed the formation of two resonance modes that differ by their
symmetry with respect to the exchange of adjacent copper oxide
layers \cite{pail1,woo06}. The original resonance mode observed in
the bilayer cuprate possesses an odd (o) symmetry while the new one
exhibits an even (e) symmetry. The frequency of the even mode is
larger while its intensity is smaller than that of the odd mode.
Moreover, while the doping dependence of the odd mode is
non-monotonic and roughly follows $\Omega_{res}^{o} \sim 5k_B T_c$
\cite{he}, the frequency of the even mode increases monotonically
with decreasing doping \cite{private,capogna}.

The splitting in energy between odd and even resonances has been
analyzed theoretically in the past within the random phase
approximation (RPA) \cite{millis_monien}, and has been attributed to
the difference in the strength of the residual interaction leading
to the bound state. The larger the interaction, the more the
resonance is shifted downwards from the lower edge of the
particle-hole ($ph$) continuum.  How such a difference in the
interaction can easily be seen in the $t-J$ model, where the
interactions in the even and odd spin channels are given by
\begin{equation}
J_{o,e}({\bf q})=J_\parallel ({\bf q}) \pm J_\perp \label{eq:Joe}
\end{equation}
with $J_\parallel,J_\perp>0$ being the in-plane and out-of-plane
exchange interaction, respectively. Thus $J_{o}>J_{e}$, and the odd
resonance occurs at a lower energy than the even one. Moreover,
since the even mode lies closer to the $ph$ continuum its intensity
is lower than that of the odd one. These two theoretical results
\cite{millis_monien,brinck,yamase} are in good agreement with the
experimental observations \cite{pail1,woo06,private,capogna}.

In this article, we address three issues which have not yet been
considered in earlier studies on the spin resonance in bilayer
systems. First, we argue that the difference between the even and
odd modes comes from two factors. One is the difference in the
interaction, which was taken into account in earlier studies,
another is the difference in the free-fermion susceptibilities of
the even and odd channels which so far has been neglected. We show
that the two factors are generally comparable to each other and
depend on the same combination of parameters. Numerically, the
difference in the interactions leads to a larger splitting between
the even and odd resonances than the difference between the even and
odd free-fermion susceptibilities. Second, we extend our previous
analysis of the odd resonance's dispersion \cite{prev} to the even
channel, and show that the even resonance mode also disperses
downwards at deviations from ${\bf Q}$. Moreover, we show that the
downward dispersion of the even mode is more parabolic than that of
the odd channel. Third, we demonstrate that there exists a second
branch of the even resonance, similar to the recently observed
second branch (the $Q^*$-mode \cite{prev}) of the odd resonance,
\cite{pai,hay}. We show, following the approach of Ref.\cite{prev},
that in the even channel, this second branch is much narrower in
energy than in the odd one. These results suggest further
experimental test that may finally resolve the long-standing
question regarding the origin of the resonance peak.

Finally, we analyze the doping dependence of the even and odd
resonances. In the overdoped region, both modes decrease due to a
decreasing superconducting gap. In the opposite limit of zero doping
even and odd resonances very likely evolve into the acoustic and
optical spin wave modes of the bilayer Heisenberg antiferromagnet.
We show, however, that, while plausible, the crossover from one
regime to the other cannot be obtained within a simple RPA scheme
chiefly because of the incorrect doping dependence of the
free-fermion susceptibilities:  the real part of both, the even and
odd susceptibility decreases with decreasing doping at half-filling
\cite{onufrieva}. This behavior is a direct consequence of the fact
that the even susceptibility diverges at the van-Hove singularity,
and the odd susceptibility possesses a maximum near the van-Hove
point.

The rest of the paper is organized as follows. In Sec.~\ref{Q} we
introduce our theoretical model and discuss the origin of the
splitting between the even and odd resonance at ${\bf Q}=(\pi,\pi)$.
In Sec.~\ref{dispersion} we present the dispersion of the two
resonances away from ${\bf Q}$ and show that a $Q^*$-mode also
arises in the even channel. In Sec.~\ref{DopingDependence} we
discuss the doping dependence of the resonances. Finally, in
Sec.~\ref{summary} we summarize our results and conclusions.

\section{Even and odd resonances at ${\bf Q}=(\pi,\pi)$}
\label{Q}

\begin{figure}[b]
\epsfig{file=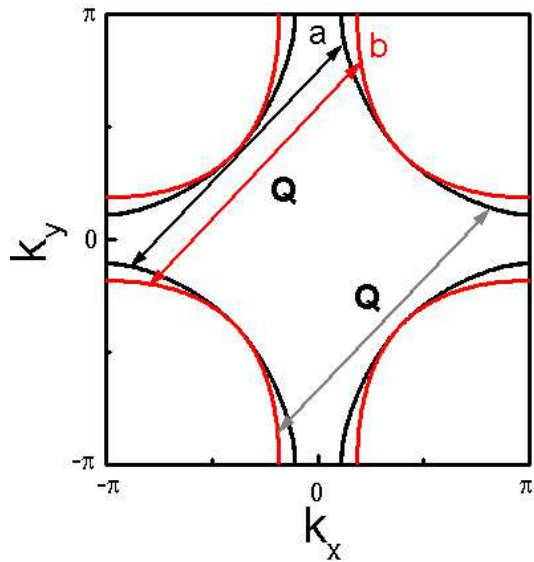,width=7.cm} \caption{(color online)
Calculated Fermi surface(FS) for the bilayered cuprates as obtained
from Eq.(\protect\ref{eq:1}). The arrows indicates the transition
between bonding-bonding ({\it bb}), antibonding-antibonding ({\it
aa}) and antibonding-bonding ({\it ab, ba}) states for
antiferromagnetic wave vector {\bf Q}$ = (\pi,\pi)$.} \label{fig1}
\end{figure}

The coupling between two CuO$_2$-planes in a unit cell of YBCO is
described by the interlayer hopping matrix element $t_{\perp}({\bf
k})=\frac{1}{4} t_{\perp}\left( \cos k_x - \cos k_y \right)^2$, Ref.
\onlinecite{tperp}. This coupling leads to the formation of bonding
({\it b}) and antibonding ({\it a}) energy bands whose dispersion
are given by
\begin{eqnarray}
\varepsilon_{\bf k}^{a,b} & = & -2t\left( \cos k_x +\cos k_y \right) +
4t'\cos k_x \cos k_y \nonumber \\
&&\pm \frac{1}{4} t_{\perp}\left( \cos k_x - \cos k_y \right)^2 -
\mu \quad, \label{eq:1}
\end{eqnarray}
with $t=250$meV, $t'/t=0.4$, $t_{\perp}/t=0.2$, and $\mu$ being the
chemical potential (these parameters provide a good fit to the Fermi
surface of the bilayered Bi$_2$Sr$_2$CaCu$_2$O$_{8+\delta}$ (BSCCO)
\cite{kordyuk}). The resulting Fermi surfaces for the bonding and
antibonding bands are shown in Fig.~\ref{fig1}.
%
%

The bonding and antibonding creation and annihilation operators are
related to the fermionic operators, $c_{1,2}$ in the two layers via
\begin{equation}
c_a = \frac{c_1 + c_2}{\sqrt{2}},~~c_b = \frac{c_1 - c_2}{\sqrt{2}},
\end{equation}
It is also convenient to introduce even and odd components of the
spins at site ${\bf i}$, which are given by
\begin{eqnarray}
{\bf S}_e({\bf i}) &=& \frac{{\bf S}_1({\bf i}) + {\bf S}_2({\bf
i})}{2} \nonumber \\
&=& \frac{1}{2} \left(c^\dagger_{a,\alpha}({\bf i}) {\bf
\sigma}_{\alpha,\beta} c_{a,\beta}({\bf i})
+ c^\dagger_{b,\alpha} ({\bf i}) {\bf \sigma}_{\alpha,\beta} c_{b,\beta} ({\bf i}) \right) , \nonumber \\
\nonumber \\
{\bf S}_o({\bf i}) &=& \frac{{\bf S}_1 ({\bf i}) - {\bf S}_2({\bf i})}{2} \nonumber \\
&=& \frac{1}{2} \left(c^\dagger_{a,\alpha}({\bf i}) {\bf
\sigma}_{\alpha,\beta} c_{b,\beta}({\bf i}) +
c^\dagger_{b,\alpha}({\bf i}) {\bf \sigma}_{\alpha,\beta}
c_{a,\beta}({\bf i}) \right) . \nonumber\\
\end{eqnarray}
The experimentally measured susceptibility is related to the even
and odd susceptibilities, $\chi^{e} = <S_e S_e>$ and $\chi^o = <S_o
S_o>$ via \cite{ben1}
\begin{eqnarray}
\chi({\bf q}, \omega) & = & \chi^{e}({\bf q}, \omega)\cos^2 \frac{q_z d}{2}
+ \chi^{o}({\bf q}, \omega)\sin^2 \frac{q_z d}{2},
\label{eq:2}
\end{eqnarray}
where $d$ is the separation between the layers. For non-interacting
electrons, the susceptibilities in the even and odd channels are
given by $\chi_0^e=\chi_0^{aa}+\chi_0^{bb}$ and
$\chi_0^o=\chi_0^{ab}+\chi_0^{ba}$, respectively, where
$\chi_0^{aa}$ and $\chi_0^{bb}$ represent intraband  particle-hole
excitations, and $\chi_0^{ab}$ and $\chi_0^{ba}$ represent interband
excitations.  The free-fermion susceptibilities in the
superconducting state at $T=0$ are given by \cite{excit,com5}
\begin{widetext}
\begin{equation}
\chi^{ij}_0 ({\bf q}, \omega) = \frac{1}{4} \sum_{\bf k} \left( 1 -
\frac{\varepsilon^i_{\bf k}\varepsilon^j_{\bf k+q}+\Delta^i_{\bf k}
\Delta^j_{\bf k+q}}{E^i_{\bf k} E^j_{\bf k+q}} \right) \left(
\frac{1}{\omega + E^j_{\bf k+q}+E^i_{\bf k}  + i\Gamma} -
\frac{1}{\omega -E^j_{\bf k+q}-E^i_{\bf k} + i\Gamma}\right)
\label{eq:chiab}
\end{equation}
\end{widetext}
with $i,j=a,b$, $E^i_{\bf k}=\sqrt{\left(\varepsilon^i_{\bf
k}\right)^2+\left(\Delta^i_{\bf k} \right)^2 }$, and $\Delta^i_{\bf
k}$ is the superconducting gap in the bonding ($i=b$) and
antibonding ($i=a$) band. In the following we assume that the
pairing part of the Hamiltonian is symmetric with respect to the
bilayers and given by
\begin{eqnarray}
{\cal H}_{pp} &=& \sum_{\bf k} \Delta ({\bf k})
\left(c^\dagger_{1,\uparrow} ({\bf k}) c^\dagger_{1,\downarrow}
(-{\bf k}) +
c^\dagger_{2,\uparrow} ({\bf k}) c^\dagger_{2,\downarrow} (-{\bf k}) \right. \nonumber \\
& & \left. + h.c. \right)\nonumber \\
& = & \sum_{\bf k} \Delta ({\bf k}) \left(c^\dagger_{a,\uparrow}
({\bf k}) c^\dagger_{a,\downarrow} (-{\bf k}) +
c^\dagger_{b,\uparrow} ({\bf
k}) c^\dagger_{b,\downarrow} (-{\bf k}) \right. \nonumber \\
& & \left. +h.c. \right) \quad,
\end{eqnarray}
where $\Delta ({\bf k}) = \frac{\Delta_0}{2} (\cos k_x - \cos k_y)$.
It then follows that the pairing gap is the same for bonding and
antibonding bands, implying $\Delta^a_{\bf k}=\Delta^b_{\bf
k}=\Delta_{\bf k}$. However, the respective Fermi surfaces in both
bands are located at different momenta ${\bf k}$ in the Brillouin
zone. In order to obtain the full $\chi^{e,o}$, we use the
random-phase approximation (RPA). Within RPA the even and odd parts
of the full spin susceptibility are given by
\begin{equation}
\chi_{RPA}^{\alpha}({\bf q},\omega)=\frac{\chi^\alpha_0({\bf
q},\omega)}{1-g_\alpha({\bf q}) \chi^\alpha_0({\bf q},\omega)} \quad,
\label{eq:chiRPA}
\end{equation}
where $\alpha=o,e$ and $g_{e,o}({\bf q})$ are the fermionic interaction
vertices in the even and odd channels.

We first consider the spin susceptibility at momenta close to ${\bf
Q} = (\pi,\pi)$. The dominant contribution to the susceptibilities
comes from fermions near the hot spots, where both ${\bf k}$ and
${\bf k+Q}$ are close to the Fermi surface. In a $d_{x^2-y^2}$-wave
superconductor with the above $\Delta ({\bf k})$ one has $\Delta
({\bf k+Q}) = - \Delta ({\bf k})$. As a consequence,
Im$\chi^{e,o}_0$ exhibits discontinuities due the opening of the
superconducting gap\cite{Morr00}. For the odd susceptibility,
Im$\chi_0^{ab}$ and Im$\chi_0^{ba}$ exhibit a single discontinuity
at $\Omega^{ab}_c({\bf Q})=|\Delta^a_{\bf k}| + |\Delta^b_{\bf
k+Q}|$, where ${\bf k}$ is chosen such that $\varepsilon_a ({\bf k})
= \varepsilon_b ({\bf k+Q}) =0$ (see Fig.\ref{fig1}). Below this
frequency, Im$\chi_0^{ab} = 0$ (at $T=0$).  At the same time,
Im$\chi^{e}_0$ possesses two discontinuities located at
$\Omega^{aa}_c({\bf Q}) =|\Delta^a_{\bf k}| + |\Delta^a_{\bf k+Q}|$
and $\Omega^{bb}_c({\bf Q}) = |\Delta^b_{\bf k}| + |\Delta^b_{\bf
k+Q}|$, where ${ \bf k}$ is again chosen such that both fermions are
at the Fermi surface (see Fig.~\ref{fig1}). Im$\chi^{e}_0$ is zero
below the lower discontinuity, and jumps between two finite values
at the higher discontinuity. Analyzing Eq. (\ref{eq:chiab}), we find
$\Omega^{bb}_c({\bf Q})<\Omega^{ab}_c({\bf Q})<\Omega^{aa}_c({\bf
Q})$.

Hence, in the even and odd channel, the susceptibility at low
frequencies is purely real and, according to Eq.(\ref{eq:chiab}),
one finds that the bare $\chi^{\alpha}_0({\bf Q},\omega)$
($\alpha=a,b$) behaves as
\begin{equation}
\chi^{\alpha}_0({\bf Q},\omega) = \chi^{\alpha}_0({\bf Q},0) -
A_{\alpha} f(\omega/\Omega_{ij}) \ , \label{eq.:chiexp}
\end{equation}
where $A_\alpha>0$, and $f(x) \propto x^2$ at small $x$, and $f(x)
\sim |\log (1-x)|$ near $x =1$.  Substituting this result in
Eq.(\ref{eq:chiRPA}), one finds that since $f(x)$ changes between
$0$ and $\infty$ when $x$ changes between $0$ and $1$, the
susceptibilities in both the odd and even channel develop resonances
below the thresholds of the particle-hole continuum, at frequencies
$\Omega_{e,o}$ where $1 = g_{e,o} \chi^{e,o}_0({\bf
Q},\Omega_{e,o})$.

As mentioned above, there are two reasons why the resonances in the
even and odd channels occur at different frequencies. One is that
the even and odd free-fermion susceptibilities, $\chi^{e,o}_0({\bf
Q},\omega)$ are different, another one is that the interactions are
different in the even and odd channels. The difference between the
resonance frequencies in the even and odd channels due to the
difference in $\chi^{e,o}_0({\bf Q},\omega)$ arises predominantly
from the fact that the (dimensionless) magnetic correlation lengths
$\xi_{e,o}^{-2} = 1 - g_{e,o} \chi^{e,o}_0 ({\bf Q}, 0)$ are
different in both channels already in the normal state. Additional
differences between $\chi^{e,o}_0({\bf Q},0)$ which arise in the
superconducting state scale as $\Delta_0/E_F$, and are hence small
and can be neglected. Assuming that the difference between the even
and the odd resonance is small, and that the resonance frequencies
are sufficiently low, such that $f(x)$ in Eq.(\ref{eq.:chiexp})
scales as $x^2$, we find after some simple algebra that
\begin{equation}
\frac{\Omega_e - \Omega_o}{\Omega_o} =  \frac{\xi^{-1}_e - \xi^{-1}_o}{\xi^{-1}_o}
\end{equation}
The r.h.s.~of the above equation is in turn related to the
difference in the normal state static $\chi$ via
\begin{equation}
\frac{\xi^{-1}_e - \xi^{-1}_o}{\xi^{-1}_o} = \frac{g}{2} \xi^2_o
\left(\chi^{ab}_0 + \chi^{ba}_0 - \chi^{aa}_0 - \chi^{bb}_0\right) \
. \label{ch_1}
\end{equation}
The dominant contributions to the r.h.s. of Eq.(\ref{ch_1}) come
from fermions in hot regions near $(0,\pi)$ and $(\pi,0)$, for which
the $t_{\perp}$ term in the dispersion,  Eq.(\ref{eq:1}) reduces to
$\pm t_{\perp}$. Expanding the r.h.s. of  Eq.(\ref{ch_1}) to leading
order in $t_{\perp}$, we obtain
\begin{equation}
\frac{\xi^{-1}_e - \xi^{-1}_o}{\xi^{-1}_o} = t^2_{\perp} \frac{g
\xi^2_o}{2\pi^3} \int\frac{d \omega d^2k}{(\epsilon_k - i\omega)^2
(\epsilon_{k+q}-i\omega)^2} \label{ch_2}
\end{equation}
where $\epsilon_k$ is the in-plane dispersion (i.e., Eq.(\ref{eq:1})
with $t_{\perp}=0$).  Linearizing $\varepsilon_{\bf k}$ and
$\varepsilon_{\bf k+Q}$ in the hot regions  as $v_F (k_x +
k_y)/\sqrt{2}$ and $v_F( k_x - k_y)/\sqrt{2}$, respectively,
substituting this expansion into the susceptibilities, and
performing the integration, we obtain
\begin{equation}
\frac{\xi^{-1}_e - \xi^{-1}_o}{\xi^{-1}_o} = t^2_{\perp} \frac{8 g
\xi^2_o}{\pi^2 v^3_F k_{max}} \label{ch_3}
\end{equation}
where $k_{max} \sim k_F$ is the upper limit of the integration over
momentum. Observe that the r.h.s. of Eq.(\ref{ch_3}) is {\it
positive}, implying that because of the difference in the even and
odd susceptibilities, the resonance in the even channel occurs at a
larger frequency than the resonance in the odd channel. To estimate
the strength of the effect, we use $v_F k_F \sim 0.5 eV \sim 2 t$,
$g = U/4$, where $U$ us the Hubbard interaction (in the RPA,
$mU/2\pi \approx 1$), set $k_{max} = k_F$, and define $J_{\perp} = 4
t^2_{\perp}/U$, $J = 4t^2/U$.  We then have
 \begin{equation}
\frac{\xi^{-1}_e - \xi^{-1}_o}{\xi^{-1}_o} = \frac{\xi_o^2}{4\pi}
\frac{J_\perp}{J} \label{eq:ch_4}
\end{equation}

The second source for the difference between $\Omega_e$ and
$\Omega_o$ is the difference in the interaction strength between the
two channels. As mentioned above,  within the $t-J$ model, the two
interactions are given by $J_{o,e}=J_\parallel ({\bf q}) \pm
J_{\perp}$. At ${\bf q=Q}$, this effect alone leads to
\begin{equation}
\frac{\xi^{-1}_e - \xi^{-1}_o}{\xi^{-1}_o} = \frac{J_\perp}{J} \ ,
\label{eq:ch_5}
\end{equation}
where $J=J_\parallel ({\bf q=Q})$. We see that both effects
described by Eqs.(\ref{eq:ch_4}) and (\ref{eq:ch_5}) are in fact of
the same order, {\it and both lead to a larger $\Omega_e$ compared
to $\Omega_o$}.  Moreover, the effect of the $t_{\perp}$ dependence
of the interaction is larger, at least near optimal doping, where
$\xi_o \sim 1$. However, with decreasing doping, and hence
increasing $\xi_o$, the role of the difference in the even and odd
free-fermion susceptibilities may become more dominant.

\begin{figure}[h]
\epsfig{file=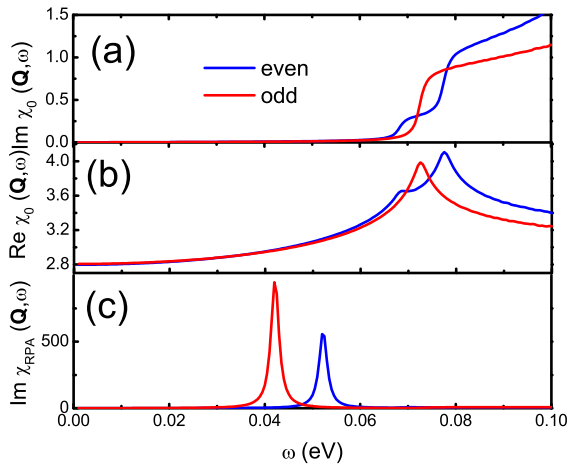,width=8.5cm} \caption{(color online) (a)
Im$\chi^{e,o}_0$,  (b) Re$\chi^{e,o}_0$ and (c) Im$\chi^{e,o}_{RPA}$
as a function of frequency at the antiferromagnetic wave vector
${\bf Q}=(\pi,\pi)$.} \label{fig2}
\end{figure}

In Fig.~\ref{fig2}, we present the results for the bare and full
susceptibilities at optimal doping, corresponding to $\mu=-1.195 t$,
that were obtained from a numerical evaluation of
Eqs.(\ref{eq:chiab}) and (\ref{eq:chiRPA}).
We see that Re$\chi_0$ in even and odd channels are almost identical
below $2\Delta_0$, {\it i.e.}  the difference in the
susceptibilities is too small to give rise to an observable
difference between $\Omega_e$ and $\Omega_o$. This agrees with our
analytic treatment. Hence, the difference between $\Omega_e$ and
$\Omega_o$ arises from the difference in the effective interactions
$g_e$ and $g_o$. To reproduce the experimentally measured frequency
splitting between both resonances at ${\bf Q}$, and  the dispersion
of the two modes (see below), we use
\begin{equation}
g_{o,e}({\bf q})=J_0 \left\{1-0.1
\left[\cos(q_x)+\cos(q_y)\right]\right\} \pm 0.027 J_0 .
\label{eq:J}
\end{equation}
According to Eq.(\ref{eq:Joe}), the first (second) term on the
r.h.s. of the above equation can be interpreted as arising from the
in-plane (out-of-plane) exchange interaction $J_\parallel ({\bf q})$
($J_\perp$).

We present the RPA susceptibilities $\chi_{RPA}^{e,o}$ at ${\bf Q}$
in Fig.~\ref{fig2}(c). We see that both even and odd
susceptibilities show resonance behavior. By construction, the
resonance in the even channel occurs at a larger frequency than the
odd resonance. Accordingly, the intensity of the even resonance is
smaller, which agrees well with the experimental
observations~\cite{pail1}.
\begin{figure}[h]
\epsfig{file=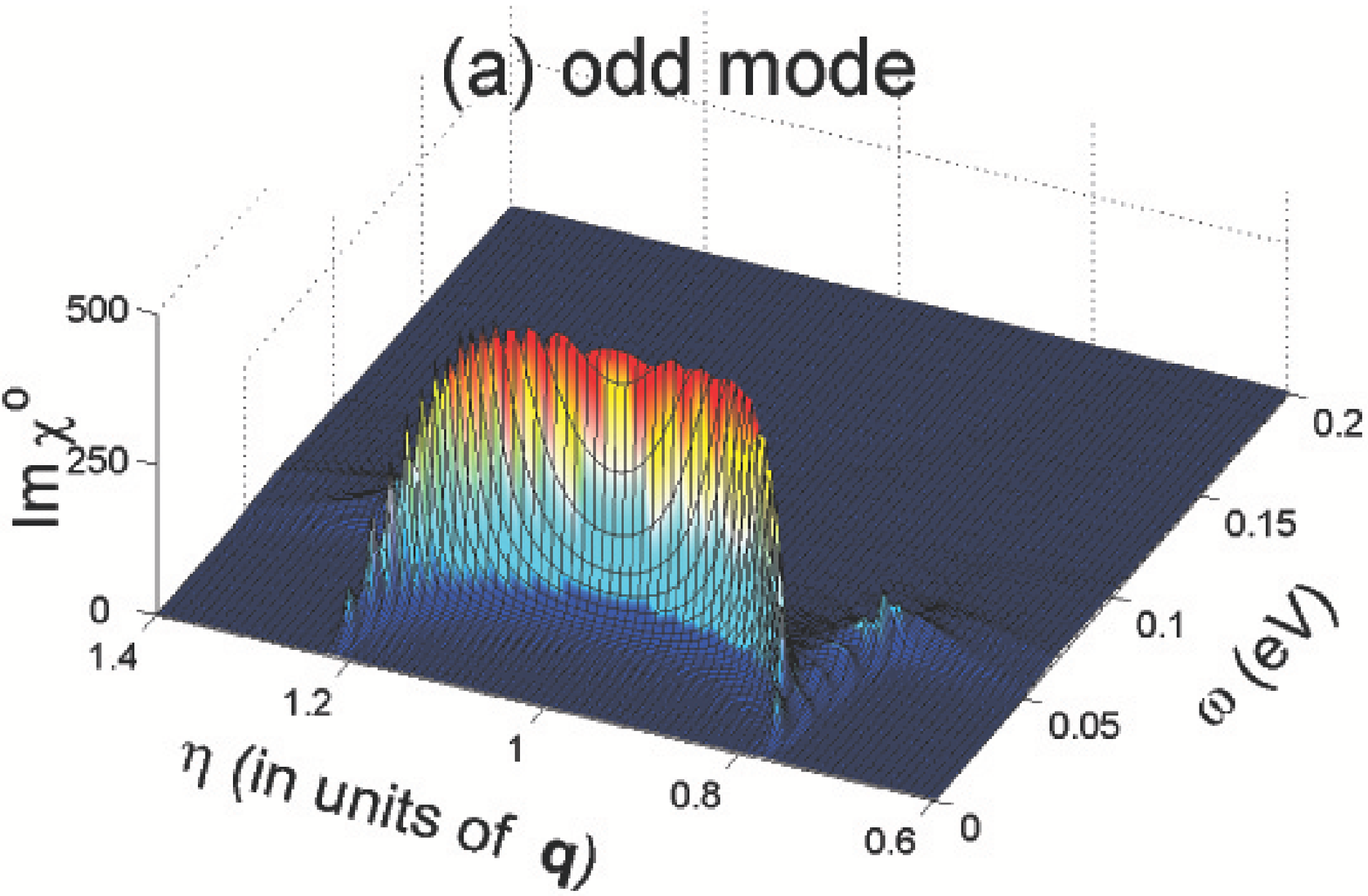,width=8cm}
\epsfig{file=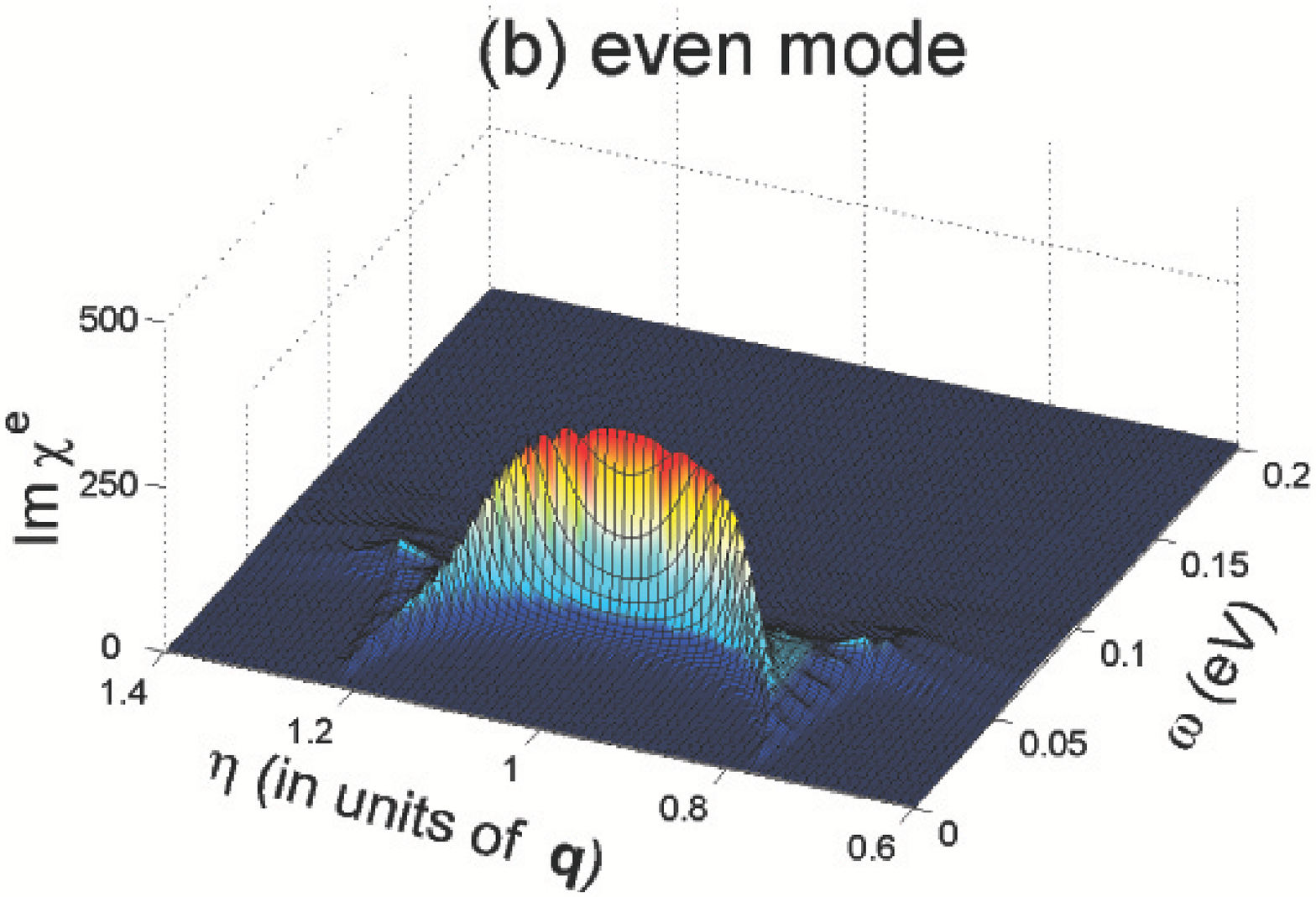,width=8cm}
 \caption{(color online) RPA results for
magnetic excitations in a bilayered $d_{x^2-y^2}$ superconductor at
optimal doping. Calculated Im$\chi^o$(a) and Im$\chi^e$(b) obtained
from Eq.~(\ref{eq:chiRPA}) as a function of momentum along ${\bf
q}=\eta (\pi,\pi)$) and frequency in the SC state.} \label{dispers}
\end{figure}

\section{The dispersion of the resonance peak}
\label{dispersion}

We next consider the dispersion of the even and the odd resonances
and present in Fig.~\ref{dispers} a plot of Im$\chi^{e,o}_{RPA}
({\bf q}, \Omega)$ at optimal doping along ${\bf q}=\eta (\pi,\pi)$
as a function of frequency.  The momentum dependence of the odd
mode's frequency and intensity, shown in Fig.~\ref{dispers}(a), is
quite similar to that of the resonance mode in the single-layer
model\cite{prev}. In particular, away from {\bf Q} three
discontinuities in Im$\chi^o_0$ emerge, corresponding to scattering
channels with momenta ${\bf q}$, $(2\pi,0)-{\bf q}$, and
$(2\pi,2\pi)-{\bf q}$. The first momentum corresponds to a direct
transition, while the last two momenta describe scattering processes
involving Umklapp scattering \cite{prev}. As discussed before, the
resonance can occur only at frequencies below the lowest
discontinuity in Im$\chi^o_0$ \cite{excit,prev}. Since the
superconducting gap decreases towards the diagonal of the Brillouin
Zone (BZ), the resonance dispersion follows the momentum dependence
of the {\it ph} continuum, forming a parabolic-like shape
\cite{excit,prev}. Upon reaching ${\bf Q}_0 \approx (0.8, 0.8) \pi$
corresponding to the wave vector connecting the nodal points of the
superconducting gap on the Fermi surface, the spin gap vanishes, and
no resonance is possible. For even smaller {\bf q} one finds that
another resonance branch emerges, the so called {\it Q}$^*$ mode,
arising from an umklapp transition \cite{prev}.

In contrast, the even part of the spin susceptibility exhibits six
discontinuities in Im$\chi_0^e$ away from ${\bf Q}=(\pi,\pi)$.
Intraband scattering within the bonding and antibonding bands each
gives rise to three of these discontinuities. Similarly to the odd
susceptibility, we find that a genuine resonance occurs only below
the lowest discontinuity in Im$\chi_0^{bb}$ due to the direct
transition with momentum ${\bf q}$. This transition is again
responsible for the parabolic-like shape of the even mode's
dispersion, as shown in Fig.~\ref{dispers}(b). However, we find that
the intensity of the even resonance falls off much faster as one
moves away from ${\bf Q}$ than that of the odd one. Since the
superconducting gap and the splitting of the Fermi surfaces is zero
along the diagonal of the BZ, the position of the so-called ``silent
band" is the same for the odd and for the even channel. Thus, both
resonances merge together at ${\bf Q}_0 \approx (0.8, 0.8) \pi$ (see
also Fig.\ref{fig:dopdep}(c)). Similar to the resonance in the odd
channel, we find that for momenta smaller than ${\bf Q}_0$,  an
umklapp transition leads to the formation of a ${\it Q}^*$ mode in
the even channel. However, its energy range is much smaller than
that of the odd ${\it Q}^*$ mode due to the proximity to the $ph$
continuum. For the odd as well as the even resonance mode we find
that while the intensity of the $Q$ mode (i.e., the mode originating
at ${\bf Q}$) is largest along ${\bf q}= (\pi, \eta \pi)$ and ${\bf
q}=(\eta \pi, \pi)$, the $Q^*$ mode has its largest intensity along
the diagonal direction, i.e., along ${\bf q}=\eta(\pi,\pi)$ and
${\bf q}=[(2-\eta) \pi, \eta \pi]$. As previously discussed
\cite{prev}, this rotation of the intensity pattern by $45^\circ$
reflects the qualitative difference in the origin of the first and
second resonances.

\section{Doping dependence of the even and odd resonances}
\label{DopingDependence}

\begin{figure}[h]
\epsfig{file=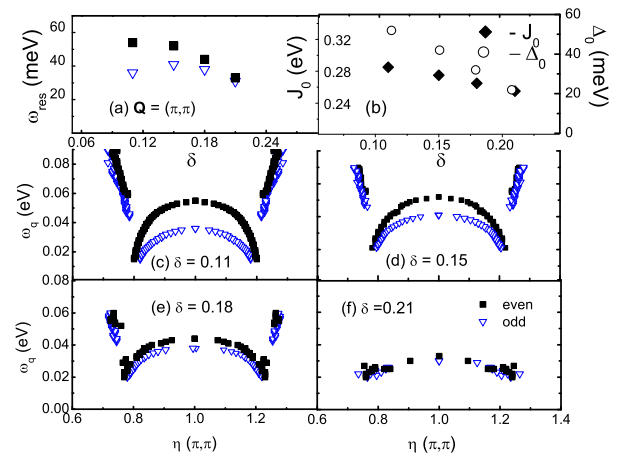,width=8.9cm} \caption{(color online) (a)
Doping dependence of (a) the resonance frequency at ${\bf Q}$ in the
odd and even channel, and (b) the superconducting gap $\Delta_0$ and
$J_0$. (c)-(f) Dispersion of the even and odd modes for various
doping concentrations in the (c) underdoped, (d) optimally doped,
and (e) and (f) overdoped regime.} \label{fig:dopdep}
\end{figure}

Next, we consider the doping dependence of the resonance modes in
the odd and even channels. In order to describe the doping
dependence, it is necessary to know that of the superconducting gap
as well as that of $g_{o,e}({\bf q})$. The doping dependence of the
superconducting gap, which is shown in Fig.~\ref{fig:dopdep}(b), is
taken from recent ARPES experiments \cite{eschrnorm} which suggest
that the superconducting gap increases by about $10 \div 20 \%$
going from the optimally doped to the underdoped cuprates.
In order to describe the doping dependence of $g_{o,e}({\bf q})$, we
leave the momentum dependence of $g_{o,e}({\bf q})$ unchanged, and
only change the overall prefactor, $J_0$, in Eq.(\ref{eq:J}), as a
function of doping, by fitting the frequency of the resonance in the
odd channel. The doping dependence of $J_0$ is also shown in
Fig.~\ref{fig:dopdep}(b). We find that this procedure provides a
satisfactory fit to the experimentally measured dispersion of both
resonance modes over a considerable range of doping.

In Fig.~\ref{fig:dopdep}(a) we present the doping dependence of the
resonance in the even and odd channels at ${\bf Q}=(\pi,\pi)$. As
expected from the discussions above, we find that with increasing
doping, the energy splitting between both modes decreases, and for
$\delta=0.21$ is only about $\Delta \omega_{res} \approx 1$ meV at
${\bf Q}$, while for $\delta=0.15$ one has $\Delta \omega_{res}
\approx 12$ meV. This decrease in the splitting is observed over the
entire dispersion of the resonance modes in the even and odd
channels, which we present in Figs.~\ref{fig:dopdep}(c)-(f) for
several different doping levels. In addition, we find that the
dispersion of the even mode exhibits a continuous downshift with
increasing doping, while that of the odd mode first shifts upwards
with increasing doping in the underdoped systems, but shift
downwards in the overdoped regime. In order to understand this
qualitative difference between the underdoped and overdoped region,
we note that in general, the doping dependence of the resonance
modes is determined by that of the superconducting gap (which in
turn determines that of the {\it ph} continuum) as well as that of
$g_{o,e}({\bf q})$. While a decrease of the superconducting gap, and
hence a downward shift in frequency of the $ph$ continuum leads to a
downward shift of the resonances, a decrease of $g_{o,e}({\bf q})$,
in contrast, leads to an upward shift of the modes' dispersion.

Since the dispersion of the even resonance is located in frequency
close to the $ph$ continuum, and Re$\chi_0^{e}$ varies strongly in
the vicinity of the $ph$ continuum due to its logarithmic
singularity, it follows that the dispersion of the even resonance is
rather insensitive to changes in $g_{e}({\bf q})$. As a result, the
doping dependence of the even resonance is predominantly determined
by that of the $ph$ continuum, exhibiting a continuous downward
shift in energy with increasing doping. In contrast, in the
underdoped regime, the energy difference between the $ph$ continuum
and the odd mode's dispersion is rather large, and Re$\chi_0^{o}$
varies only weakly around the resonance frequency. As a result, the
resonance frequency is very sensitive to changes in $g_{o}({\bf
q})$. Therefore, it is the decrease in $g_{o}({\bf q})$ with
increasing doping (and not the decrease in the superconducting gap)
that determines the doping dependence of the odd mode's dispersion
and leads to its upward shift in energy in the underdoped regime.
Around optimal doping, the odd mode's dispersion has become
sufficiently close to the $ph$ continuum, that the mode's further
doping dependence is now determined by that of the $ph$ continuum,
and not any longer by that of $g_{o}({\bf q})$, similar to the case
of the even mode. Hence, the two opposite effects arising from a
decrease of the superconducting gap and that of $g_{o}({\bf q})$
lead to the qualitatively different doping dependence of the odd
mode's dispersion in the underdoped and overdoped regime. Note that
with increasing doping, and the resulting downward shift of the $ph$
continuum, the momentum range over which the $Q^*$-mode can be
observed, decreases.

\begin{figure}[h]
\epsfig{file=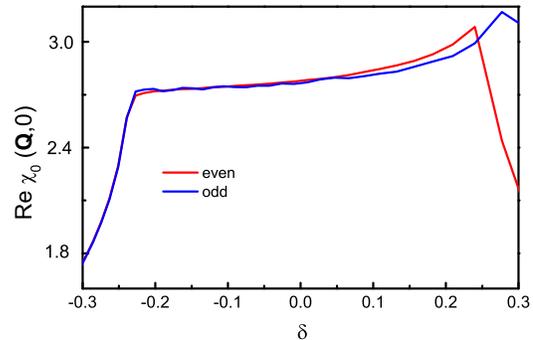,width=7.9cm} \caption{(color online)
Re$\chi^{e,o}_0({\bf Q}, 0)$ as a function of doping concentration
in the normal state.} \label{fig:rechi0dop}
\end{figure}

Finally, we briefly discuss the doping dependence of
$\chi^{e,o}({\bf Q},\omega=0)$. If indeed, the suggested above, the
odd and even resonance are transformed into the acoustic and optical
branches of the spin wave dispersion in the antiferromagnetically
ordered phase, one would expect that $\chi^o_0({\bf Q},\omega=0)$
increases with decreasing doping. As a result, one would see a
downward shift in the odd mode's dispersion even for a doping
independent $J_0$. One finds, however, that the doping dependence of
$\chi^o_0({\bf Q},0)$, which is obtained from Eq.(\ref{eq:chiab}) by
simply changing the chemical potential, $\mu$, defies this
expectation. This is shown in Fig.\ref{fig:rechi0dop}, where we
present the doping dependence of $\chi^{o,e}_0({\bf Q},0)$.
Note that the even susceptibility possesses two logarithmic
divergences as a function of doping which occur when either the
bonding or antibonding Fermi surfaces touches the van Hove (vH)
points ($\pm \pi,0$) and ($0, \pm \pi$) and undergo a topological
transition from a hole-like to and electron-like Fermi surface.
These transitions occur at a doping level of $x \approx 0.23$ for
the antibonding band and at $x \approx 0.55$ for the bonding band
(not shown). In contrast, the odd susceptibility, which arises from
scattering transitions between the bonding and antibonding bands
does not exhibit a logarithmic divergence, but is simply enhanced
and exhibits a finite maximum. If we define the minimum distance (in
momentum space) of the bonding and antibonding Fermi surfaces to the
vH point $(0,\pi)$ by $k_a(\mu)$ and $k_b(\mu)$, respectively, then
Re$\chi^{o}_0({\bf Q},0)$ exhibits a maximum at that doping level
for which the smaller of $k_a(\mu)$ and $k_b(\mu)$ possesses a
maximum. Defining $k_{min}(\mu)=\min [k_a(\mu),k_b(\mu)]$ one finds
\begin{equation}
{\rm Re}\chi^{o}({\bf Q},0) \sim const. + \frac{1}{2\pi t}
\arcsin\left[k^2_{min}(\mu) \frac{t}{t_\perp}\right]
\end{equation}
Note that for doping levels below that at which the van Hove
singularity in Re$\chi^{e}_0({\bf Q},0)$ or the maximum in
Re$\chi^{o}_0({\bf Q},0)$ occurs, the susceptibilities decrease
monotonically with decreasing doping, as shown in
Fig.~\ref{fig:rechi0dop}. This doping dependence clearly reflects a
shortcoming of the weak-coupling approach used above, which fails to
capture the strong correlation effects that are not only responsible
for the occurrence of antiferromagnetism, but are very likely also
the key ingredients in the explanation of the pseudo-gap region in
the underdoped cuprates. It is interesting to note in this context
that recent studies of the doping dependence of $\chi_0({\bf Q},0)$
for a single layer system within the FLEX approach find that the vH
singularity is eliminated by interaction effects, and that starting
from the overdoped region $\chi_0({\bf Q},0)$ increases
monotonically with decreasing doping \cite{Ereminpc}. This
shortcoming of the approach used above is effectively compensated by
a phenomenologically introduced doping dependence of $g^{e,o}$ which
increases with decreasing doping. This phenomenological approach,
however, does not allows us to fully explain the doping dependence
of the resonant excitations in the underdoped cuprates. In
particular, it leaves open the question how the downward dispersion
of the resonance mode observed in the optimally doped cuprates is
transformed into the upward dispersion of the acoustic spin-wave
branch.

\section{Summary}
\label{summary}

In this study, we have investigated the form of magnetic resonance
excitations in the even and odd spin channel of the bilayer cuprates
in the superconducting state. We obtain a number of new results
suggesting further experimental test that may finally resolve the
long-standing question concerning the origin of the resonance peak.
First, we show that the energy splitting between the even and odd
resonances arises not only from a different interaction strength in
both channels, but also from a the difference in the free-fermion
susceptibilities in the even and odd channels. Both effects scale as
$\sim J_\perp/J$ and lead to a frequency for the even resonance that
is larger than that of the odd resonance. However, at least at
optimal doping, the numerical prefactors are such that the energy
splitting is dominated by the difference in the interaction strength
and not by the difference in the free-fermion susceptibilities.
Since the latter scales with $\xi_o^2$, the relative importance of
these two effects might change in the underdoped cuprates. In
agreement with previous results \cite{millis_monien,brinck,yamase}
we also find that the intensity of the even resonance is weaker than
that of the odd resonance. Second, we computed the dispersion of the
even resonance and showed that the even resonance also disperses
downward as one moves away from ${\bf Q}=(\pi,\pi)$. Moreover, we
demonstrated that the downward dispersion of the even mode is more
parabolic than that of the odd channel. Third, we showed that there
exists a second branch of the even resonance, similar to the
recently observed second branch (the $Q^*$-mode \cite{prev}) of the
odd resonance, \cite{pai,hay}. We find, however, that in the even
channel, this second branch is much narrower in energy than in the
odd one. Fourth, we studied the doping dependence of the both
resonance modes, and find that that of the even mode is determined
by the downward shift of the $ph$ continuum with increasing doping.
In contrast, the upward shift in frequency of the odd resonance in
the underdoped cuprates is determined by the decrease in $g_o$ with
increasing doping, while in the overdoped regime, the odd resonance
follows the doping dependence of the $ph$ continuum. Our results
demonstrate that the structure of magnetic excitations in the
superconducting state of the bilayered cuprates is dominated by the
topology of the Fermi surface, the interaction strength in the even
and odd channel, and the $d_{x^2-y^2}$-wave symmetry of the
superconducting gap.

Finally, we note that the experimental situation has recently been
complicated by the report that an even resonance exists at
incommensurate wave vectors only \cite{woo06}. This result
contradicts earlier studies which have found that the even resonance
exhibits the largest intensity at ${\bf Q}=(\pi,\pi)$\cite{private}.
The origin of this experimental discrepancy is currently unclear.

We thank Y. Sidis, Ph.~Bourges, P. Dai, and I.A. Larionov  for
helpful discussions. The present work is supported by the DAAD
Collaborative German-US research grant No. D/05/50420. D.K.M.
acknowledges financial support by the Alexander von Humboldt
Foundation, the National Science Foundation under Grant No.
DMR-0513415 and the U.S. Department of Energy under Award No.
DE-FG02-05ER46225. A.C. acknowledges support from NSF DMR 0604406.


\begin{thebibliography}{99}

\bibitem{YBCO}
J. Rossat-Mignod, L. P. Regnault, C. Vettier, P. Bourges, P. Burlet,
J. Bossy, J. Y. Henry and G. Lapertot, Physica C {\bf 185-189}, 86
(1991); H.F. Fong, P. Bourges, Y. Sidis, L.P. Regnault, J. Bossy, A.
Ivanov, D.L. Milius, I.A. Aksay, and B. Keimer,
 Phys. Rev. B {\bf 61}, 14773 (2000); P. Dai, H.
A. Mook, R. D. Hunt, F. Dogan, Phys. Rev. B {\bf 63}, 054525 (2001).

\bibitem{Bi}  H.F. Fong, P. Bourges, Y. Sidis, L.P. Regnault, A. Ivanov,
G.D. Gu, N. Koshizuka, and B. Keimer, Nature (London) {\bf 398}, 588
(1999).

\bibitem{Tl} H. He, P. Bourges, Y. Sidis, C. Ulrich, L.P. Regnault, S. Pailhes, N.S.
Berzigiarova, N.N. Kolesnikov, and B. Keimer, Science {\bf 295},
1045 (2002).

\bibitem{excit} H.F. Fong, B. Keimer, P.W. Anderson, D. Reznik, F. Dogan, and I.A. Aksay
 Phys. Rev. Lett. {\bf 75}, 316 (1995);
Ar. Abanov and A.V. Chubukov, Phys. Rev. Lett. {\bf 83}, 1652
(1999); J. Brinckmann and P. A. Lee, Phys. Rev. Lett. {\bf 82}, 2915
(1999); Y.-J. Kao, Q. Si, and K. Levin, Phys. Rev. B {\bf 61},
11898(R) (2000); F. Onufrieva and P. Pfeuty, Phys. Rev. B {\bf 65},
054515 (2002); D. Manske, I. Eremin, and K.~H. Bennemann, Phys. Rev.
B {\bf 63}, 054517 (2001); M.R. Norman, Phys. Rev. B {\bf 61}, 14751
(2000); {\it ibid} {\bf 63}, 092509 (2001);  A. Chubukov,  B. Janko
and O. Tchernyshov, Phys. Rev. B {\bf 63}, 180507(R) (2001); I.
Sega, P. Prelovek, and J. Bonca, Phys. Rev. B {\bf 68}, 054524
(2003).

\bibitem{other} L. Yin, S. Chakravarty, and P.W. Anderson, Phys. Rev.
Lett. {\bf 78}, 3559 (1997); E. Demler and S.C. Zhang, Phys. Rev.
Lett. {\bf 75}, 4126 (1995); D.K. Morr and D. Pines, Phys. Rev.
Lett. {\bf 81}, 1086 (1998).

\bibitem{other1} M. Vojta and T. Ulbricht, Phys. Rev. Lett. {\bf 93}, 127002
(2004); G.S. Uhrig,  K.P. Schmidt, and M. Gr\"uninger, Phys. Rev.
Lett. {\bf 93}, 267003 (2004); F. Kr\"uger and S. Scheidl, Phys.
Rev. B {\bf 70}, 064421 (2004); G.S. Uhrig, K.P. Schmidt, and M.
Gr\"uninger, J. Phys. Soc. Jpn. (Suppl.) {\bf 74}, 86 (2005); M.
Vojta, T. Vojta, and R.K. Kaul, Phys. Rev. Lett. {\bf 97}, 097001
(2006).

\bibitem{other2} G. Seibold, and J. Lorenzana
Phys. Rev. B {\bf 73}, 144515 (2006); {\it ibid.} Phys. Rev. Lett.
{\bf 94}, 107006 (2005).

\bibitem{pail1} S. Pailhes, Y. Sidis, P. Bourges, C. Ulrich, V. Hinkov, L.P. Regnault, A. Ivanov,
B. Liang, C.T. Lin, C. Bernhard, and B. Keimer, Phys. Rev. Lett.
{\bf 91}, 237002 (2003).

\bibitem{woo06} H. Woo, P. Dai, S.M. Hayden, H.A. Mook, T. Dahm, D.J. Scalapino, T.G. Perring,
and F. Dogan, Nature Physics (London) {\bf 2}, 600 (2006).

\bibitem{he} H.F. He, Y. Sidis, P. Bourges, G.D. Gu, A. Ivanov, N. Koshizuka, B. Liang, C.T. Lin,
L.P. Regnault, E. Schoenherr, and B. Keimer, Phys. Rev. Lett. {\bf
86}, 1610 (2001).

\bibitem{private} S. Pailhes, C. Ulrich, B. Fauque, V. Hinkov, Y. Sidis, A. Ivanov, C.T. Lin, B. Keimer, and
P. Bourges, Phys. Rev. Lett. {\bf 96}, 257001 (2006).

\bibitem{capogna} L. Capogna, B. Fauque, Y. Sidis, C. Ulrich, Ph. Bourges, S. Pailhes, A. Ivanov,
J.L. Tallon, B. Liang, C.T. Lin, A.I. Rykov, and B. Keimer,
cond-mat/0610869 (unpublished).

\bibitem{millis_monien}  A.J. Millis and H. Monien, Phys. Rev. B {\bf 54}, 16172 (1996).


\bibitem{brinck} J. Brinckmann and P.A. Lee, Phys. Rev. B {\bf
65}, 014502 (2001); T. Li, Phys. Rev. B {\bf 64}, 012503 (2001).

\bibitem{yamase} H. Yamase, and W. Metzner, Phys. Rev. B {\bf 73},
214517 (2006).


\bibitem{prev} I. Eremin, D.K. Morr, A.V. Chubukov, K.H. Bennemann, and M.R. Norman,
Phys. Rev. Lett. {\bf 94}, 047001 (2005).

\bibitem{pai}  S. Pailhes,  Y. Sidis, P. Bourges, V. Hinkov, A. Ivanov, C. Ulrich, L.P. Regnault, and
B. Keimer, Phys. Rev. Lett. {\bf 93}, 167001 (2004); D. Reznik, P.
Bourges, L. Pintschovius, Y. Endoh, Y. Sidis, T. Masui, and S.
Tajima, Phys. Rev.  Lett. {\bf 93}, 207003 (2004).

\bibitem{hay} S.M. Hayden, H.A. Mook, P.C. Dai, T.G. Perring, and F. Dogan, Nature (London) {\bf
429}, 531 (2004).

\bibitem{onufrieva}
F. Onufrieva and P. Pfeuty, Phys. Rev. Lett {\bf 95}, 207003 (2005).

\bibitem{tperp} O.K. Anderson, A.I. Liechtenstein, O. Jepson, and F. Paulsen, J. Phys. Chem. Solids {\bf 56},
1573 (1995).

\bibitem{kordyuk} A.A. Kordyuk,  S.V. Borisenko, M. Knupfer, and J. Fink, Phys. Rev. B {\bf 67},
064504 (2003).

\bibitem{ben1} Note, $\chi({\bf q}, \omega)$ results from the Green's function
$\langle\langle S^+_{\bf q}| S^-_{\bf -q} \rangle \rangle$, where
the spin operators are $S^+_{\bf q}=S_{1{\bf q_{||}}}^+
e^{iq_z\frac{d}{2}}+ S_{2{\bf q_{||}}}^+ e^{-iq_z\frac{d}{2}}$ and
$S^+_1=S_e^+ + S_o^+$, $S^+_2=S_e^+ - S_o^+$. This transformation
yields Eq.(\protect\ref{eq:2}).


\bibitem{com5} For the numerical calculation of $\chi_0$,
we employed $\delta=2$ meV in the analytic continuation of the
Greens functions $i \omega_n \rightarrow \omega+i\delta$.


\bibitem{Morr00} D.K.~Morr and D. Pines, Phys. Rev. B \textbf{62}, 15177
(2000); {\it ibid.} Phys.~Rev.~B \textbf{61}, R6483 (2000).


\bibitem{eschrnorm} M. Eschrig, and M.R. Norman, Phys. Rev. B {\bf 67},
144503 (2003).


\bibitem{Ereminpc} T. Dahm, and I. Eremin, Phys. Rev. Lett. {\bf 97}, (2006) (in press).


\end{thebibliography}
\end{document}